\documentclass[aps,showpacs,twocolumn,amsmath,amssym,groupedaddress]{revtex4}
\usepackage{dcolumn}
\usepackage{amsmath,amsfonts}
\usepackage{color,graphicx}
\usepackage{dcolumn}
\usepackage{bm}
\begin{document}
\title[Acoustically forced charged bubble dynamics]{Effect of charge on the dynamics of an acoustically forced bubble}
\author{Thotreithem Hongray$^1$}
\author{B. Ashok$^2$}
\altaffiliation[Author to whom correspondence should be addressed. ]{Electronic mail:  bashok1@gmail.com, bashok@iiitb.ac.in\\}
\author{J. Balakrishnan$^{1}$}
\altaffiliation[Current address: ]{School of Natural Sciences \& Engineering, National Institute of Advanced Studies (N.I.A.S.), Indian Institute of Science Campus, Bangalore - 560012, India.} 
\affiliation{$^1$School of Physics, University of Hyderabad, Central Univ. PO, Gachi Bowli, Hyderabad 500 046, India.}
\affiliation{$^2$International Institute of Information Technology, Bangalore (I.I.I.T.-B),\\ 26/C, Electronics City, Hosur Road, Bangalore 560 100, India.}
\begin{abstract}  
The effect of charge on the dynamics of a gas bubble undergoing forced oscillations 
in a liquid due to incidence of an ultrasonic wave is theoretically investigated. 
The limiting values of the possible charge a bubble may physically carry are obtained.
The presence of charge influences the regime in which the bubble's radial 
oscillations fall.
The extremal compressive and expansive dimensions of the bubble are also 
studied as a function of the amplitude of the driving pressure. 
It is shown that the limiting value of the bubble charge is dictated both by 
the minimal value reachable of the bubble radius as well as the amplitude of the 
driving ultrasound pressure wave. 
A non-dimensional ratio ${\zeta}$ is defined that is a comparative measure of 
the extremal values the bubble can expand or contract to and find the existence 
of an unstable regime for ${\zeta}$ as a function of the driving pressure amplitude, $P_s$. 
This unstable regime is gradually suppressed with increasing bubble size. The 
Blake and the upper transient pressure thresholds for the system are then discussed.\\
\end{abstract}

\maketitle

\section{Introduction}
The study of bubble dynamics and cavitation has a long and interesting history in the scientific 
literature. One of the earliest works was that of Lord Rayleigh~\cite{rayleigh} in his study of cavitation phenomena, motivated by the need to understand and minimize the damage to ships' propellors due to cavitation  
(the low pressure on the surface of the propeller blades causes the liquid in contact with the surface to spontaneously form unstable bubble clouds which  often self-organize into dendritic structures\cite{neppiras}. These bubble clouds implode with enormous force resulting in serious damages to the propeller).\\
Cavitation, bubble formation and dynamics are present in different instances and situations. 
Analyses and studies of the phenomena have been motivated by and have explained very distinct 
natural, practical phenomena. 
Small amplitude oscillations of a gas bubble in a liquid were studied by Minnaert~\cite{minnaert}. 
His work showed the important contribution of radial oscillations of entrailed air bubbles in the 
sound heard from running water.\\ 
In nature, the snapping shrimp uses rapid closure of its claws to generate cavitating bubbles which 
stun its prey~\cite{versluis}. Bubble formation and kinetics contribute to fluid flow in biological 
systems in blood and cells in the micron scale~\cite{marmot}. In the presence of incident ultrasound 
waves, bubbles can also enhance the rate of chemical reactions~\cite{suslick, masonbook}.
Bubble formation and cavitation can be recreated under controlled conditions in the laboratory by 
subjecting a liquid in a container to a standing ultrasonic wave, setting up a pressure field within 
the liquid. When the driving pressure amplitude of the sonic field becomes larger than the ambient 
pressure, the pressure in the liquid becomes negative. When this pressure exceeds the vapour pressure 
of the liquid, local evaporation is caused. The liquid `breaks' up forming tiny micron size 
cavitation bubble clouds, which implode violently within a very short period of time. Frenzel 
and Schultes demonstrated that these cavitation clouds emitted low intensity visible 
light~\cite{frenzel}. This phenomenon, wherein light is emitted by a gas bubble in a liquid 
due to its rapid  expansion and collapse when ultrasound is incident on it, is known as 
sonoluminescence. This has also contributed to an extensive study of bubble dynamics in the context 
of sonoluminescence. Several other studies have followed in the literature, including those of Gaitan 
and others~\cite{gaitan1,gaitan2,gaitan3,barberPutterman}. The radial oscillations of gas bubbles in 
a liquid that are caused due to incident acoustic waves cannot be described trivially. These are a 
type of driven nonlinear oscillations that can greatly depend on initial conditions and can be chaotic 
in nature. This has been shown conclusively (see, for example,~\cite{brennerRMP,lauterborn, smereka, 
lauterborn2, parlitz} and references therein). Reviews may be found in, for example, 
~\cite{masonbook,brennerRMP,brennenbook}.\\ 

The presence of electric charge on bubbles in fluids has been reported in the experimental literature.
~\cite{mctaggart,alty,alty2,akulichev,shiran}.\\ 
In this paper we use the Rayleigh-Plesset equation modified by Parlitz, {\it et al.}, further 
modified to take into account the presence of charge on the bubble, and study the effects of charge, 
driving frequency and amplitude of the ultrasonic driving field on the dynamics of the bubble, 
with the specific heat ratio being taken as 5/3 in our calculations, consistent with the gas 
in the bubble being a monatomic ideal, inert gas under adiabatic conditions. 
In Section II, we describe the model used; the natural frequency of oscillation 
of the bubble is obtained and the phase plots show that the maximal radial velocity 
for the charged bubble greatly exceeds that of the neutral bubble. In Section III 
the Blake radius and threshold for the charged bubble are calculated. This is followed by 
a discussion of Rayleigh collapse and the influence of driving pressure amplitude in Section IV. 
We show that for a bubble of given ambient radius $R_0$, there exists a cut-off value for the 
maximum charge it can carry, which is dictated by the van der Waals hard-core radius. 
In Section V we introduce a new dimensionless ratio $\zeta$ as a measure of the maximum and 
minimum radius a bubble can achieve under acoustic driving. This when plotted as a function 
of pressure amplitude clearly captures the positions 
of the Blake threshold and the upper critical transient pressure threshold for acoustic cavitation, 
and their distinct dependence on the pressure amplitude, driving frequency and bubble charge. 
Bubbles that are present in various systems can carry a non-zero charge. Hence, a proper understanding 
of their behavior and dependence on pressure, driving frequency, ambient radius and other parameters, 
is essential. This is the subject of investigation of this paper. The results obtained in our work 
therefore become very useful in the context of medical uses of acoustic cavitation and ultrasound 
for diagnostic and therapeutic purposes.

\section{The Model}
We shall briefly describe the model and the set of equations being used to describe the 
radial oscillations of gas bubbles in liquids. 
Various models have investigated distinct, disparate limits of bubble-collapse. 
The original work of Rayleigh assumed the surrounding liquid to be inviscid and 
incompressibile~\cite{rayleigh}. Plesset and others ~\cite{plesset1,plesset2,noltingkNeppiras} 
have included viscosity, surface tension etc. Keller and Kolodner used the same 
expression but with a modification for accounting for acoustic radiation by 
the bubble by considering the liquid as slightly compressible~\cite{kellerKolodner}.
Keller and Miksis further included all these modifications -- of viscosity, 
surface tension, the incident sound wave and acoustic radiation, in one model 
to obtain a modified equation~\cite{kellerMiksis}. 

The model we use to describe the system is based on the earlier equations of Prosperetti, Parlitz 
and Keller and Miksis and others~\cite{parlitz,kellerMiksis,plessetProsperetti,prosperetti-etal} 
and which are modifications of the Rayleigh-Plesset equations~\cite{rayleigh,plesset1,plesset2}. 

The system we consider is that of a bubble suspended in a liquid and that has some net charge. 
The presence of charge on a bubble is not speculative. As mentioned earlier, electric charge 
has been found on bubbles under acoustic forcing. Various reasons could cause charge to be present 
on the bubble, e.g., the migration of ionic charges in the liquid onto the bubble surface, 
although the exact mechanism has been debated.  See, for example, the work of Alty~\cite{alty,alty2} 
and Akulichev~\cite{akulichev}. 
That gas bubbles in water can carry a charge has been clearly demonstrated long back~\cite{alty,alty2}. 
Indeed, it had been shown earlier~\cite{mctaggart} that air and other gas bubbles in water 
become negatively charged. More recently, this has been shown by Shiran and Watmough~\cite{shiran}. 
Therein, bubbles placed in an electric field are clearly demonstrated to veer towards one of 
the electrodes.
\begin{figure}[]
\includegraphics[height=8.5cm,width=6cm,angle=0]{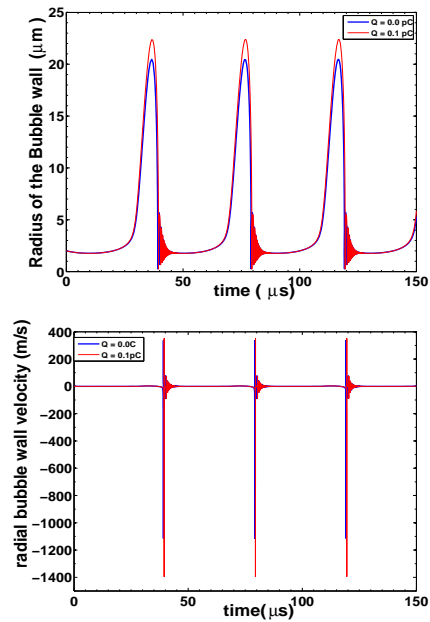}
\caption{Plots of a bubble's radius (above) \& radial velocity (below) as a function of time 
($P_s = 1.35 P_0$, $R_0 = 2 \mu m$, $\nu = 25$ kHz). 
The curve for the charged bubble (with charge 0.1 pC) reaches a larger magnitude (for both 
radius and velocity plots) than for the neutral bubble. (Color Online).}
\label{radiusplotR2} 
\end{figure} 
Other work on the presence of charge on bubbles in different situations include that done by 
Bunkin and Bunkin~\cite{bunkin}.\\ 
The study of the dynamics of charged gas bubbles in fluids is important because such systems 
find various applications, including in medicine -- a widely used one being in the medical 
application of ultrasound. 
While it has been overlooked in several models, various investigations of charged bubbles have 
appeared in the literature in different contexts. These include, for example,~\cite{zharov}. 
Their work is however of very limited scope since they consider the specific case for which 
the polytropic index $\Gamma$ takes the value 4/3 and for this case it turns out that the terms 
containing the charge mutually cancel out in the Rayleigh-Plesset equation. \\ 

We take $\Gamma=5/3$ consistent with taking the heat transfer across the bubble to be an 
adiabatic process. For the sake of simplicity, we assume the charge to be strictly limited to, 
and uniformly distributed on, the surface of the bubble.\\ 
The surface charge density of the bubble thus changes as the bubble expands and contracts 
along with the acoustic forcing being externally applied to it. 
As mentioned above, we use a modified form of the Rayleigh-Plesset equation, one introduced by 
Parlitz, et al.~\cite{parlitz}, which is equivalent in first order in ($1/c$) to that of Keller 
and Miksis~\cite{kellerMiksis} ($c$ being the velocity of sound) and which includes, approximately, 
the sound radiation which is the most significant contribution to damping at higher amplitudes 
of bubble excitation. In the presence of charge, this equation describing the dynamical evolution 
of the bubble radius $R$ in time gets further modified.  

Considering the charged bubble as a non-conducting charged shell with a constant charge $Q$, 
the electrostatic pressure can be calculated as has been done by Akulichev, Atchley and 
others~\cite{akulichev,atchley,bunkin,zharov}. 
The presence of charge $Q$ (or surface charge density $S$) causes the inclusion of an 
electrostatic pressure term $2\pi S^2/\epsilon = Q^2/(8\pi\epsilon R^4)$ into the equation 
(see e.g.,~\cite{zharov}, also~\cite{ashok}): 
\begin{eqnarray}
&&\left[\left(1-\frac{\dot{R}}{c}\right)R+\frac{4\eta}{c\rho}\right]\ddot{R}=
\frac{1}{\rho}\left(P_0 -P_v+\frac{2\sigma}{R_0}-\frac{Q^2}{8\pi\epsilon R_0^4}\right)\nonumber\\
&&\times \left(\frac{R_0}{R}\right)^{3\Gamma}\left(1+\frac{\dot{R}}{c}(1-3\Gamma)\right) 
-\frac{\dot{R}^2}{2}\left(3-\frac{\dot{R}}{c}\right) \nonumber\\
&&+\frac{Q^2}{8\pi\rho\epsilon R^4}\left(1-\frac{3\dot{R}}{c}\right)-\frac{2\sigma}{\rho R}
-\frac{4\eta}{\rho}\left(\frac{\dot{R}}{R}\right)\nonumber\\
&&-\frac{1}{\rho}(P_0-P_v+P_s\textrm{sin}(\omega t))\left(1+\frac{\dot{R}}{c}\right)
-\frac{R}{\rho c}P_s\omega \textrm{cos}(\omega t)\label{cke}\nonumber\\
\end{eqnarray} 
where $R_0$ is the ambient bubble radius, $P_0$ is the static pressure of the, $P_s$ and 
$\omega = 2\pi\nu$ denote respectively the 
amplitude and angular frequency of the driving sonic field, $P_v = 2.34$kPa is the vapour pressure 
of the gas, $\sigma$, $\rho$ and $\eta$ denote respectively the surface tension, density and 
viscosity of the liquid surrounding the bubble. $c$ is the velocity of sound in the liquid. 
In this work, for the purpose of the numerical results reported, we consider water to be the liquid, 
with $\rho = 998 kg/m^3$, $\eta = 10^{-3} Ns/m^2$, $c = 1500 m/s$, $P_0=101kPa$, $\sigma=0.0725 N/m$, 
$\epsilon = 85\epsilon_0$ where $\epsilon_0$ is the permittivity of vacuum.\\ 
The presence of charge $Q$ modifies the influence of surface tension, reducing its effective value 
(the effective surface tension changes from $\sigma$ for the uncharged case to $\sigma 
- Q^2/(16\pi \epsilon R_0^3)$) and induces several interesting changes to the dynamics of 
bubble oscillations. \\
When a pressure wave is incident on a bubble in a liquid, the difference in pressure can cause 
expansion and rapid collapse of the bubble, this being followed immediately after by further, 
smaller oscillations which are termed as afterbounces. This entire sequence of a maximal expansion 
of the bubble followed by collapse and afterbounces, is repeated in each cycle when we have sinusoidal 
forcing by ultrasound. When a bubble is in the sonoluminescent regime, light emission occurs shortly 
after the bubble's violent contraction to a minimal radius, before the onset of aftebounces and 
repetition of this sequence. Details of the phenomenon of sonoluminescence may be found in, for 
example, the review articles~\cite{brennerRMP,barber}. 
We recast eqn.(1) in dimensionless form for ease of evaluation by redefining the radius $R$ through 
$r = R/R_0$ and time $t$ through $\tau = \omega t$, $r$ and $\tau$ being the new dimensionless radius 
and time variables.  Using the static pressure $P_0$ as the reference pressure we also define the 
dimensionless quantities , $P_{*v} = P_v/P_0$ and $P_{*s} = P_s/P_0$. Using an overdot to now denote differentiation with respect to $\tau$, the dimensionless time variable (rather than with respect 
to $t$), eqn.(1) can be written in dimensionless form as 
\begin{eqnarray}
&&\left(1-\frac{\dot{r}}{c_*}\right)r\ddot{r} + F\ddot{r}+\frac{\dot{r}^2}{2}\left(3-\frac{\dot{r}}{c_*}\right) \nonumber \\ 
& =& G\left(1-P_{*v}+M\right)\left(\frac{1}{r}\right)^{3\Gamma}\left(1+\frac{\dot{r}}{c_*}(1-3\Gamma)\right)\nonumber \\
&+&\frac{C}{r^4}\left(1-\frac{3\dot{r}}{c_*}\right)-S\frac{1}{r}-Fc_*\left(\frac{\dot{r}}{r}\right) \nonumber \\
&-&G\left(1-P_{*v}+P_{*s}\sin(\tau)\right)\left(1+\frac{\dot{r}}{c_*}\right)-G\frac{rP_{*s}}{c_*}\cos(\tau)
\label{nondimeq}\nonumber \\
\end{eqnarray}
where 
\begin{eqnarray*}
&&c_*=\frac{c}{R_0\omega}; \,\,\,\, F=\frac{4\eta}{\rho R_0 c}; \,\,\,\,  G=\frac{P_0}{R_0^2\omega^2\rho};\\
&&M=\frac{1}{P_0}\left(\frac{2\sigma}{R_0}-\frac{Q^2}{8\pi\epsilon R_0^4}\right)\\
&&C=\frac{Q^2}{8\pi\epsilon R_0^6{\omega}^2\rho};\,\,\,\,S=\frac{2\sigma}{\rho R_0^3\omega^2}
\end{eqnarray*}
are all dimensionless constants.\\
This form is used when solving the equation numerically. In what follows, we will use the dimensional 
form of the equation everywhere. It should be understood though, that data points shown in all the graphs 
have been obtained after numerical evaluation of the corresponding dimensionless quantities, followed by 
rescaling by the appropriate multiplicative factors to obtain the variables in physically realizable units.\\ 
Figure \ref{radiusplotR2} shows a plot of the radius of the bubble as a function of time, obtained by solving 
equation (\ref{nondimeq}), for charge present on, as well as absent from, the bubble surface. \\
For a given driving frequency of the pressure wave, the maximum radius attainable by the bubble 
increases with charge present. Conversely, the minimum radius achievable also reduces with increasing 
charge. These are as expected.\\
The difference in behaviour can be seen more emphatically in a plot of the bubble's radial velocity 
as a function 
of time (in Figure \ref{radiusplotR2}). In the plot shown, in the presence of charge, the maximum 
velocity increases to more than five quarters of its uncharged value. 
This gives us a picture of the dynamics consistent with the time-series of the bubble radius, namely, 
that the bubble oscillations become more violent in the presence of charge, causing the bubble to expand 
to a greater extent and contract to a lesser minimal radius 
than in the absence of any surface charge.\\
  
\begin{figure}[!h]
\includegraphics[height=5.5cm,width=6cm,angle=0]{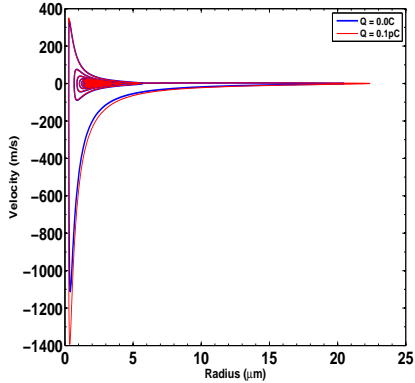}
\caption{$\dot{R}$ vs. $R$ phase plots for charged and uncharged bubble. The charged case ($Q = 0.1$ pC) 
spans larger magnitudes of velocity and radius. $R_0=2\mu m, P_s=1.35P_0, \nu=25kHz$. (Color Online).}
\label{phaseplotR2} 
\end{figure}

The phase portrait for the system is shown in Fig.{\ref{phaseplotR2}} for both a charged as well as 
uncharged bubble. 
The phase plot for the charged bubble is larger than that of the uncharged case, reaching larger 
values of the magnitudes of both radius and velocity. 
One can intuitively expect this. During the expansion part of the cycle, the charges present on the bubble 
surface move away from each other as the bubble expands, decreasing the surface charge density of the
 bubble wall. The change in the value of the maximal radius of the bubble due to the presence of charge,
 though present, is small. Due to the decrease in surface tension, we would expect the radius of the 
bubble to grow slightly more than in the case when there is no charge present. On the other hand, 
during the collapse, the surface charge density increases rapidly. It is in this regime that we would 
expect to see most clearly the effects of charge on the dynamics of the bubble. Since the maximum 
radius attained is larger, the collapse would be expected to be more  violent resulting in a higher 
peak velocity of the bubble wall at the moment of collapse and lower minimum radius attained.\\ 

The natural frequency of bubble oscillations of small amplitude may be found 
by assuming that the sound field can be introduced through a perturbation 
of amplitude $\alpha$ which is small \cite{plessetProsperetti}. Then by assuming that the bubble 
oscillates about its equilibrium radius $R_0$, one can express its radius 
at time $t$ as 
\begin{equation}
R = R_0(1 + x(t)),  
\end{equation}
where $x(t)$ is a small quantity of order $\alpha$. 
Substituting Eqn.(3) in the unforced Rayleigh - Plesset equation 
and linearising it, we obtain:
\begin{equation}
\ddot{x}  + \beta \dot{x} + \omega_0^2 x = 0, 
\end{equation}
where the damping coefficient $\beta$ and the natural frequency of the oscillator are given by 
\begin{eqnarray}
\beta &=& \frac{1}{\rho c R_0\left(1 + \frac{4\eta}{c\rho R_0}\right)}\left(4\phi/R_0^5 
+ \frac{3Q^2}{8\pi\epsilon R_0^4} + \frac{4\eta c}{R_0}\right) \nonumber \\
\omega_0^2 &=& \frac{1}{\rho R_0^2\left(1 + \frac{4\eta}{c\rho R_0}\right)}\left(5\phi/R_0^5 
-\frac{2\sigma}{R_0} + \frac{4Q^2}{8\pi\epsilon R_0^4}\right)
\end{eqnarray} 
where only terms linear in $x$ and its derivatives have been retained. In eqns.(5), the quantity 
$\phi$ defined as 
\begin{equation}
\phi / R_0^5 = ( P_0 - P_v + \frac{2\sigma}{R_0} - \frac{Q^2}{8\pi\epsilon R_0^4}) .
\end{equation}
is the equilibrium gas pressure in the bubble.\\
The solutions to the damped equation eqn.(4) represent small amplitude oscillations.
Substituting the values of the various parameters in eqns.(5) yields a natural frequency 
$\omega_0$ for a micron-sized bubble that is in the MHz range, in conformity with observations 
in the literature. 

\section{The Blake threshold}
The growth of a bubble can be determined by various threshold 
conditions~\cite{apfel}. 
One is the well-known Blake threshold for mechanical growth 
of a gas bubble. The Blake threshold corresponds to the 
minimum acoustic pressure $P_s = P_{Blake}$ exceeding which will 
result in the explosive growth of the bubble, culminating in cavitation. 
Blake threshold calculations are made under the assumption that the 
pressure fields are quasistatic in nature, and the surface tension 
dominates over viscous and inertial contributions. \\

Another threshold that is important is the transient cavitation threshold. 
This is the minimum acoustic pressure required for the forced oscillating 
bubble to collapse violently after its maximal expansion at radial velocities 
at least equalling the speed of sound.\\
In this section we calculate and discuss the Blake cavitation threshold for 
the bubble. Further discussion and results related to the upper transient 
cavitation threshold will be presented in Section V.\\

One other threshold condition is that determining the diffusion 
of a gas into the liquid causing bubble expansion, which is determined by 
factors including the natural resonance frequency of the bubble, the 
acoustic forcing frequency, the gas saturation coefficient, surface tension,  
specific heat ratio and the ambient radius of the bubble~\cite{apfel}. We do not discuss 
this threshold, known as the rectified diffusion threshold, in this work.\\
The pressure of the liquid on the outer surface of the bubble wall $p_L$ may be written down: 
\begin{equation}
p_L(R(t)) = p_i(t) - \frac{2\sigma}{R} + \frac{Q^2}{8\pi\epsilon R^4} - \frac{4\eta\dot R}{R}
\end{equation}
where the pressure inside the bubble $p_i(t)$ comprises of pressure of the gas and the vapour 
pressure $P_v$: 
\begin{equation}
p_i = ( P_0 -P_v + \frac{2\sigma}{R_0} 
+ \frac{Q^2}{8\pi\epsilon R^4})\left(\frac{R_0}{R}\right)^{3\Gamma} + P_v
\end{equation}
The net pressure on the bubble wall from the surrounding liquid is~\cite{zharov}:
\begin{eqnarray}
P &=&\Big( P_0 - P_v + \frac{2\sigma}{R_0}
- \frac{Q^2}{8\pi\epsilon R_0^4}\Big){\Big(\frac{R_0}{R}\Big)}^{3\Gamma}-\frac{2\sigma}{R} 
+ \frac{Q^2}{8\pi\epsilon R^4} \nonumber\\ 
&-&4\eta\frac{\dot R}{R} + P_v - P_{ext}
\label{eqnmin}
\end{eqnarray}
where $P_{ext} = P_0 + p(t)$, $p(t)$ being the ultrasound driving pressure.\\
For our choice of the polytropic index $\Gamma= 5/3$, the change in the bubble radius resulting 
from quasistatic changes in the pressure of the liquid $p_L$ (that is, very slow pressure changes 
of $p_L$ with inertial and viscous effects being assumed negligible during bubble expansion 
and contraction) outside the bubble may be determined from the equation:
\begin{equation}
p_L = \Big( P_0 -P_v + \frac{2\sigma}{R_0} - \frac{Q^2}{8\pi \epsilon R_0^4} \Big)\Big(\frac{R_0}{R}\Big)^5 
+ P_v - \frac{2\sigma}{R} + \frac{Q^2}{8\pi \epsilon R^4}.
\end{equation}

The charge term completely changes the liquid pressure profile: in particular for small 
bubbles, the charge term dominates over surface tension, reducing its effect.\\
  
\begin{figure}[!h]
\includegraphics[height=9.75cm,width=8.5cm,angle=0]{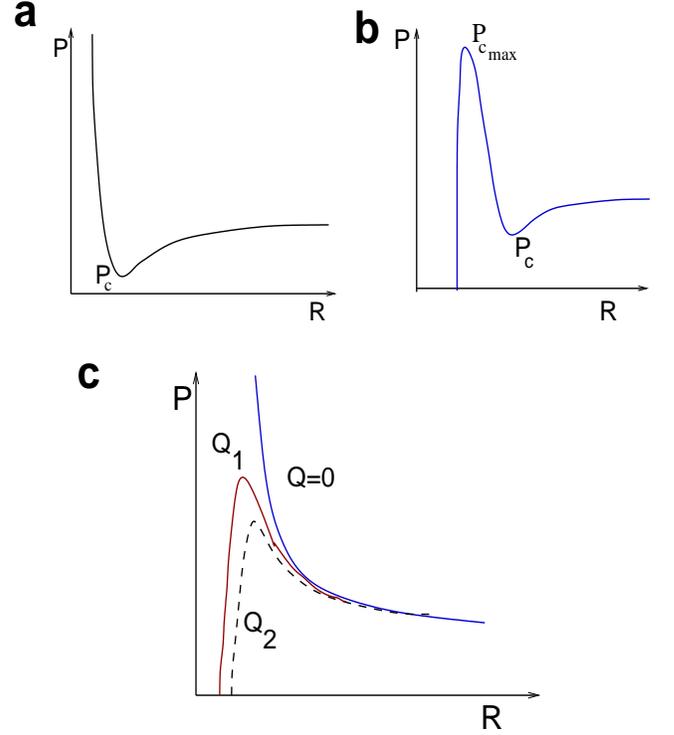}
\caption{Illustration of the behavior of eqn.(9): $P$ shown as a function of $R$ for (top left) without 
charge and (top right) with charge. The introduction of charge changes the plot dramatically. The effect 
of charge $Q$ on a bubble is shown schematically for different values of $Q$, $Q_2 > Q_1$, in the 
bottom figure. Note that our study in this paper is confined only to applied pressures in the range 
$0.4 \times 10^5 - 1.5 \times 10^5$ Pa.}
\label{effectofchargeplot} 
\end{figure}

For instance for a 5 micron bubble in water at atmospheric pressure 
($\sigma \approx $ 0.0725 N/m, $P_0 = 101$ kPa), while the surface tension contribution to the pressure 
in the bubble is roughly $2.8 \times 10^4$Pa, the effect of introducing a small charge of about 
0.415 pC would be to reduce this by half. Clearly, this has a significant effect on the radial mechanical 
stability of the bubble and the Blake threshold that determines the nature of the bubble's radial 
oscillations. 
The behavior of eqn.(9) is depicted in Figure (\ref{effectofchargeplot}) for purpose of illustration. 
In the absence of charge, there exists no equilibrium radius below a critical 
value ${P_C}$ of the pressure -- the bubble radius at this point undergoes explosive expansion. 
The presence of even a small amount of charge on the bubble surface produces a drastic change of behaviour 
-- there exists no equilibrium radius for pressures larger than a critical value ${P_C}_{max}$ ; the 
pressure region $P_C \le P \le {P_C}_{max}$ is a metastable region. 
Our study in this paper is restricted to physically realistic regimes, with applied pressures roughly 
in the range 0.4-1.5 bar.

To obtain the Blake radius for the charged bubble, we adapt the procedure for the 
uncharged case (see for example Harkin et al \cite{harkin}) to our situation. We first minimize 
equation (\ref{eqnmin}) with respect to $R$, $R>0$. This leads to the quartic equation: 
\begin{equation}
R^4 - \frac{Q^2}{4\pi\epsilon \sigma} R - \frac{5}{2\sigma}\phi = 0,
\end{equation} 
The Blake radius $R_{crit}$ is given by the real and positive root of this equation. 
We find that:
\begin{eqnarray}
R_{crit} &=& \frac{1}{2\sqrt[6]{18a}}\Bigg\{ \sqrt{ \sqrt[3]{a^2} 
- \frac{10\sqrt[3]{12}\phi}{\sigma}} \nonumber \\
&+& \frac{\frac{6Q^2}{4\pi\epsilon \sigma}\sqrt{2a} 
- \Big(\sqrt[3]{a^2} -\frac{10\sqrt[3]{12}}{\sigma}\phi\Big)^{3/2} }
{\left(\sqrt[3]{a^2} -\frac{10\sqrt[3]{12}}{\sigma}\phi\right)^{1/4} }\Bigg\}^{1/2} 
\end{eqnarray} 
where
\begin{equation} 
a = \frac{9Q^4}{(4\pi\epsilon\sigma )^2} + \sqrt 3\Big( \frac{27 Q^8}{(4\pi\epsilon\sigma)^4} 
+ \frac{4000}{\sigma^3}\phi^3 \Big)^{1/2} .
\end{equation}
The liquid pressure $p_{L_{crit}}$ corresponding to this critical value of the radius is obtained 
by substituting eqn.(9) back into eqn(4) :
\begin{equation}
p_{L_{crit}} = P_v + \frac{\phi}{R_{crit}^5} - \frac{2\sigma}{R_{crit}} 
+ \frac{Q^2}{8\pi\epsilon R_{crit}^4}
\end{equation}
The Blake threshold pressure may be obtained from the standard definition \cite{harkin} :
\begin{equation}
p_{Blake} = P_0 - p_{L_{crit}}
\end{equation}

A rough estimate of the Blake radius and threshold can be made for sub-micron sized bubbles for which 
the contributions from the static pressure is negligible in comparison with the charge - corrected  
terms. In this approximation, we find that 
\begin{equation}
R_{crit} \approx \frac{1}{2}(1 + \sqrt[3]{18})\left(\frac{Q^2}{4\pi\epsilon\sigma}\right)^{1/3} 
= 1.81\left(\frac{Q^2}{4\pi\epsilon\sigma}\right)^{1/3}
\end{equation} 
Substituting this in eqn.(8) and using eqn.(9), we find the following approximate expression for 
the Blake threshold:
\begin{eqnarray}
&&p_{Blake} = P_0 + \frac{Q^2}{8\pi\epsilon R_{crit}^4}\left(\frac{R_0}{R_{crit}} - 1\right) 
+ \frac{2\sigma}{R_{crit}} \nonumber\\
&=& P_0 + \left(\frac{4\pi\epsilon\sigma^4}{Q^2}\right)^{\frac{1}{3}} \left(-6.95 + 4.416 R_0 \left(\frac{4\pi\epsilon\sigma}{Q^2}\right)^{\frac{1}{3}} \right)
\end{eqnarray}

\section{Rayleigh collapse and the influence of driving pressure amplitude}

After the bubble attains its maximum radius $R_{max}$, it proceeds to the main collapse. Its dynamics 
during this phase is described by the Rayleigh equation: 
\begin{equation}
R\ddot{R} + \frac{3}{2}{\dot R}^2 = 0.
\end{equation}
In considering cavitation in this limit, one is essentially considering collapse of a void, 
ignoring all terms such as viscosity, surface tension, etc.
The solution for this is found to be 
\begin{equation}
R(t) = R_r \left(\frac{t_c -t}{T}\right)^{2/5},
\label{AEqn}
\end{equation}
where at $t=t_c$, the bubble collapses to a point $R=0$. 
$R_r$ is a characteristic radius, and $T$ is the time period of oscillation of the bubble. 
It may be noted that the characteristic radius 
$R_r$ in this scaling law is different from that reported in \cite{hilgenfeldt} where the polytropic 
constant was taken to be unity, corresponding to an isothermal process. Here we find an estimate for $R_r$ 
for $\Gamma = 5/3$, using a similar energy argument as in \cite{hilgenfeldt}.
Converting the potential energy $E_{pot}$ of the bubble at $R_{max}$ to kinetic energy at $R_0$ we get 
\begin{equation}
\dot{R} = -\left(\frac{2P_0}{\rho R_0^3}\right)^{1/2}\left(\frac{4\pi}{3}\right)^{\frac{\Gamma -1}{2}}R_{max}^{3\Gamma/2}
\label{Beqn}
\end{equation} 
Using Eqn.(\ref{AEqn}) in (\ref{Beqn}) we obtain 
\begin{equation}
R_r = \left(\frac{25 T^2 P_0}{2\rho} \left(\frac{4 \pi}{3}\right)^{\Gamma -1}\right)^{1/5}R_{max}^{3\Gamma/5}
\end{equation}
For $\Gamma = 5/3$, we get 
\begin{equation}
R_r =\left(\frac{25 T^2 P_0}{2\rho}(\frac{4 \pi}{3})^{\frac{2}{3}}\right)^{\frac{1}{5}}R_{max} \approx 2.006\left(\frac{P_0T^2}{\rho}\right)^{\frac{1}{5}}R_{max}
\label{Ceqn}
\end{equation}

\begin{figure}[!h]
\includegraphics[height=13cm,width=8.25cm,angle=0]{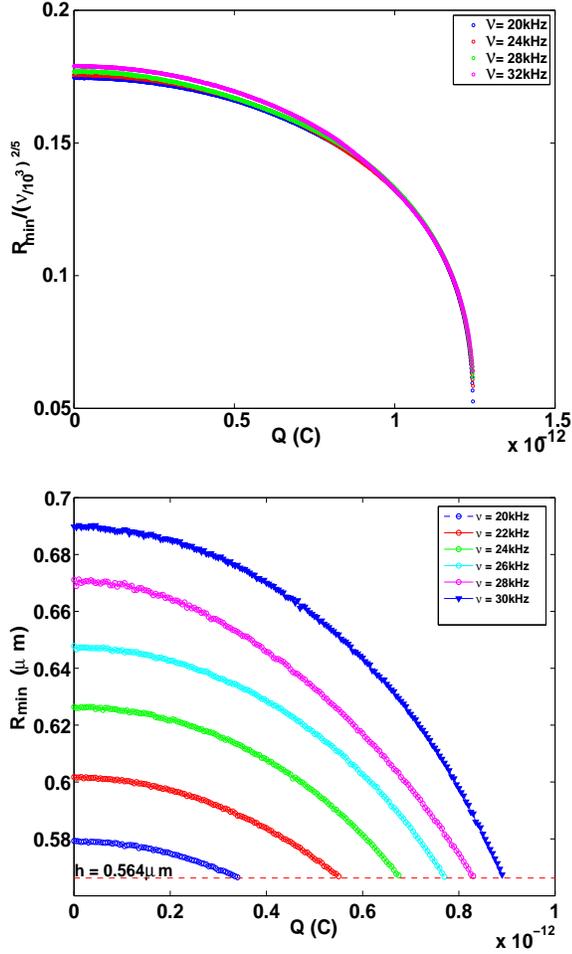}
\caption{(a)(Above): $R_{min}$ scaled by $\omega^{2/5}$, plotted as a function of charge, $Q$, for 
various driving frequencies. 
(b)(Below): Minimum radius, $R_{min}$, plotted as a function of charge, $Q$, for various driving 
frequencies ($\nu = 20, 24, 28, 32$kHz); curves for higher frequencies are at the top. The horizontal 
dotted line corresponds to the cut-off $R_{min} = h$, the van der Waals hard core radius. 
$R_0 = 5 \mu$m. (Color Online).}
\label{Rminplot} 
\end{figure}

The extremely simplified expression, Eqn.(\ref{Ceqn}), is none the less useful for making some 
physically relevant approximations. This approximation would hold best during the bubble's collapse 
to a minimum radius $R_{min}$. It would also be less inaccurate when the dimensions of the collapsing 
bubble are very small, that is, when $R_{min}$ is very small. This would tend to match more closely 
those cases where the charge on the bubble is high, so that the reduction in values of $R_{min}$ is 
correspondingly more.  
We can see that in the presence of a driving frequency 
$\omega$ for the system, $R(t)$ would show a frequency dependence $R(t) \sim \omega^{2/5}$ in the regime 
near bubble collapse, at higher driving pressure amplitudes, so that we have 
\begin{equation}
R_{min} \sim a_1\omega^{2/5}, 
\label{2by5eqn}
\end{equation}
$a_1$ being a prefactor with appropriate dimensions. \\
At higher driving frequencies, the bubble typically has larger values for its minimal radius, there 
not being sufficient time for complete collapse to occur before the expanding regime sets in. 
Increasing the charge present on the bubble enables it to reach smaller dimensions.\\ 
The minimum radius scaled by the driving frequency, through $R_{min} /\omega^{2/5}$ is shown in the 
plot of Figure (\ref{Rminplot}a). As expected from the discussion above, best agreement of 
Eqn.(\ref{2by5eqn}), as evidenced through a superposition of all the curves for different frequencies, 
is best seen at higher charge values and lower $R_{min}$ values. \\

It is to be expected however, that in reality a stable bubble can not carry an indefinite 
magnitude of charge $Q$. This can also be seen on plotting the minimum radius $R_{min}$ 
as a function of charge, where, for a given driving pressure, the minimum radius of the bubble for 
every driving frequency converges to one value of the charge. This, however, needs to be modified 
as a further physical constraint to the system exists in that bubble contraction can not also 
indiscriminately progress indefinitely. The smallest dimensions that the bubble can take, that is, 
the least value of the minimum radius $R_{min}$ reached during the bubble's compressive regime, is 
bounded by the value of the van der Waals hard core radius $h$. The value of $h$ is determined by the 
gas enclosed within the bubble. For example, for argon, $h$ has the value $h = R_0/8.86$. 
For $R_0 = 5 \mu m$, this equals a value of 0.564 $\mu m$. \\ $R_{min}$ with this constraint is shown 
in Figure (\ref{Rminplot}b).\\
There is a maximal value for the charge a bubble can carry for attainment to $R_{min} = h$ to be possible. 
This limiting value of the charge we denote by $Q_h$. This is shown in Figure (\ref{Rminplot}b), 
where the minimum radius of collapse, $R_{min}$, has been plotted  as a function of charge $Q$, for 
a bubble of initial radius $R_0 = 5\mu m$, and pressure amplitude $P_s = 1.35 P_0$. The minimal bound 
of $R_{min} = h$ has been shown by a dotted line. The point of intersection of the curve for a 
particular driving frequency with this line gives the value of $Q_h$.  

\begin{figure}[!h]
\includegraphics[height=5cm,width=9cm,angle=0]{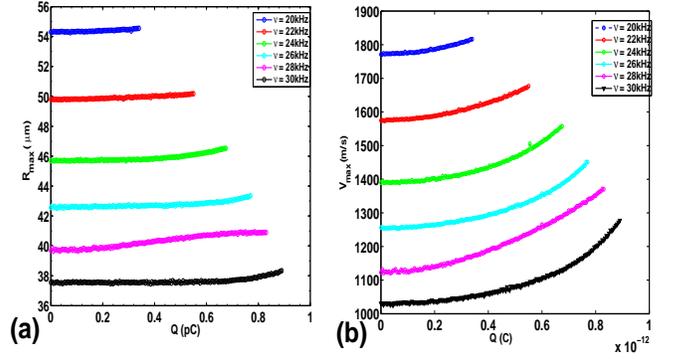}
\caption{$R_{max}$ on the left, and $v_{max}$ on the right, plotted as a function of charge, $Q$, 
for various driving frequencies ($\nu = 20,22,24,26,28,30$ kHz), lower frequency curves on the top. 
$R_0=5\mu$m, $P_s=1.35P_0$. (Color Online).}
\label{vmaxQplot}
\end{figure}

Correspondingly, there is an upper bound on the maximum radial velocity of the bubble $V_{max}$. 
The presence of charge on the bubble serves to reduce the effective magnitude of the surface tension. 
With increasing charge, the bubble is thus able to reach a smaller radius and a greater velocity.
$V_{max}$ and $R_{max}$ as a function of charge $Q$ are shown in Fig.(\ref{vmaxQplot}) 
for high driving pressure of $P_s = 1.35 P_0$. The limits to the curves in the plot are due to 
the limit in the maximal value $Q_h$ that $Q$ can take for each frequency.

Furthermore, as the frequency of the driving ultrasonic acoustic wave is increased, the maximum radius 
attained by the bubble, $R_{max}$, reduces. This is understandable by recalling that the higher the 
frequency, the shorter is the period of negative pressure shear and the bubble is driven to cavitation 
collapse in a shorter time span resulting in shorter expansion time of the bubble. This also causes 
the value of the minimum radius to become  larger with increasing frequency of the driving ultrasound wave.  
The larger the maximum radius reached, the more violent the collapse is, resulting in smaller minimum 
radius.\\

\begin{figure}
\includegraphics[height=7.5cm,width=7.5cm,angle=0]{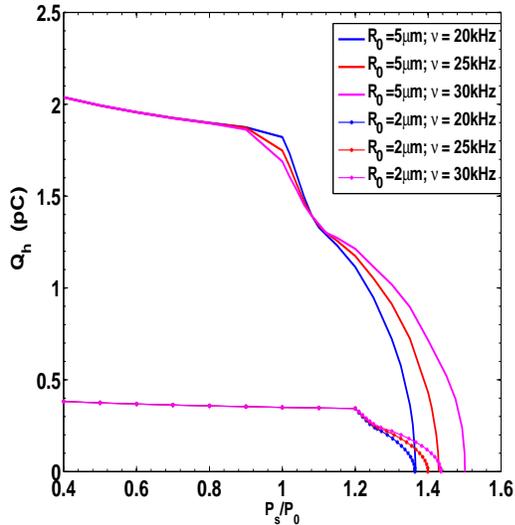}
\caption{Plot of $Q_h$ as a function of driving pressure amplitude $P_s$, for bubbles of ambient 
radius $R_0 = 2 \mu$m and $5 \mu$m. The area below each curve corresponds to the domain where 
$R_{min} \ge h$. The region above each curve corresponds to $R_{min} < h$ and is physically unreachable 
for the bubble of that particular $R_0$. $R_0=5\mu$m, $\nu = 20,25,30$kHz; $R_0=2\mu$m, 
$\nu = 20,25,30$kHz. (Color Online).}
\label{QhVsNuDiffPs}
\end{figure}

The system is very sensitive to changes in the pressure conditions.
The amplitude $P_s$ of the driving pressure determines the physically viable minimal radius attainable 
by the bubble for a given charge. The maximal, bounding value of the charge, $Q_h$, reduces 
progressively with increasing amplitude of pressure $P_s$, until beyond a critical pressure $P_m$, 
it is no longer physically possible for the bubble radius to contract to such a small value. 
In Figure (\ref{QhVsNuDiffPs}) we plot $Q_h$ as a function of $P_s$ for three different frequencies 
(20, 25 and 30 kHz). We show plots for two values of ambient bubble radius $R_0 = 2 \mu m$ and 
$R_0 = 5 \mu m$. Each curve demarcates two regions -- the space below (or to the left of) the curve 
corresponds to the physically permissible region of $R_{min} > h$. The region above (or to the right of) 
each curve corresponds to $R_{min} < h$, which can not be reached in practice by a bubble of that 
corresponding ambient radius.\\
It can be seen that the value of the critical pressure $P_m$ increases with driving frequency, and 
reduces with ambient radius $R_0$. At very low values of $P_s$, $Q_h$ becomes essentially independent 
of frequency for a given ambient radius. We denote the pressure where this frequency-independence 
first sets in (approached from above) by $P_{fi}$. It can be seen that $P_{fi}$ occurs at a lower 
value $P_{fi} = 1.12 P_0$ (with corresponding $Q_h \approx 1.3$ pC) for the larger bubble 
($R_0 = 5 \mu m$)  than  for the smaller bubble ($P_{fi} = 1.2 P_0$ for $R_0 = 2\mu$ m). Moreover 
there exists a brief crossover region for the $R_0 = 5 \mu$ m bubble where the frequency-dependence 
of $Q_h$ reverses, before true frequency-independence of $Q_h$ occurs at ${P_{fi}}_{true} = 0.9 P_0$ 
(for $P_s>P_{fi}$, $Q_h$ for higher frequencies are greater than for lower values, while 
for ${P_{fi}}_{true} < P_s < P_{fi}$, this is reversed and $Q_h$ for higher frequencies are lower 
than that for lower frequencies for the $5 \mu$m bubble).\\

To understand the effect of the driving pressure, we briefly paraphrase below 
the arguments given in \cite{hilgenfeldt} for the isothermal 
case adapting it to our adiabatic system. 

Combining the driving sound field with the static pressure \cite{plessetProsperetti}
the total external field $P_{ext}$ can be expressed as
\begin{equation}
P_{ext} = P_0(1 - \alpha \cos \omega t).
\end{equation}

Substituting this in the Rayleigh Plesset equation under quasistatic conditions:
\begin{eqnarray}
\bigg(P_0 &-& P_v + \frac{2\sigma}{R_0} 
- \frac{Q^2}{8\pi\epsilon R_0^4}\bigg)\bigg(\frac{R_0}{R}\bigg)^{3\Gamma} 
+ \frac{Q^2}{8\pi\epsilon R^4} \nonumber \\
&-&\frac{2\sigma}{R} -P_0(1-\alpha\cos\omega t) = 0 
\end{eqnarray} 
we obtain the quintic equation
\begin{eqnarray}
R^5 &-&\frac{2\sigma}{(\alpha -1)P_0}R^4 + \frac{Q^2}{8\pi\epsilon(\alpha-1)P_0}R \nonumber \\
&+& \frac{R_0^5}{\alpha-1}
\left(1 + \frac{2\sigma}{R_0P_0} - \frac{Q^2}{8\pi\epsilon R_0^4 P_0}\right) = 0
\end{eqnarray}

The behavior of the equation is completely determined by the quantity $\alpha -1$. 
For $P_{ext} > 0$, we have a a stable, single solution for $R$. 
For negative, small amplitude  $P_{ext}$ there are two solutions with that at 
lower $R$ being the stable one.\\
These two merge only at a critical value $P_{ext} = P_{Blake}$, 
$P_{Blake} < 0$. For this, $P_{gas} > P_{ext} + P_{\sigma}$ is always the 
case and leaves the equation without a solution, where $P_{gas} = \phi /R^5$ and $P_{\sigma} = 2\sigma/R$. 

Liquid pressure becoming negative, opposes the confinement effect of the surface-tension 
contribution, $P_{\sigma}$. Once the bubble is larger than a critical radius $R_c$, 
pressure balance at the bubble wall can not be maintained, and explosive growth sets in.
The bubble is unstable at this stage, and further growth leads to increased instability 
and still more expansion, and quasistatic conditions no longer hold. 

With the time period of oscillation $T = 2\pi/\omega$ being larger than the time scale 
of the bubble's oscillations, oscillations of the external pressure can  be 
considered as being quasistatic. For crossing the Blake threshold 
$P_{ext} <0$ is required so that $\alpha > 1$. 
$t=0$  gives  $P_{ext} = (\alpha -1)P_0$ to be negative.  
Thus the behavior shown by bubbles for $\alpha < 1$ will be very different from 
that for $\alpha >1$; for values of $\alpha$ less than 1, bubble oscillations 
will tend to be less violently expansive and the effect of surface tension dominate 
the dynamics.

\section{Expansion - compression ratio and the transient threshold}
\begin{figure}
\includegraphics[height=15cm,width=7.5cm,angle=0]{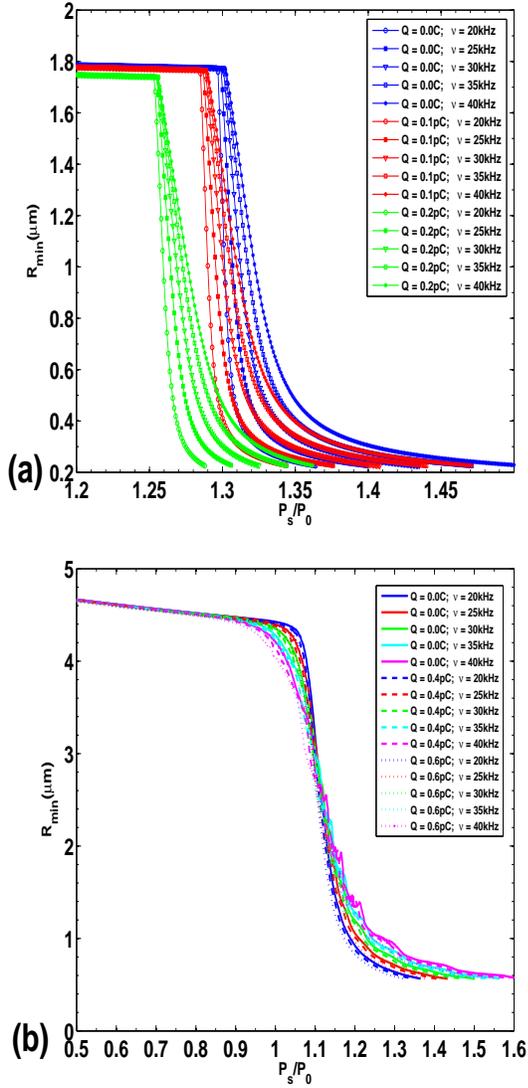}
\caption{$R_{min}$ vs $P_s$ for different values of $\omega$ ($\nu=20,25,30,35,40kHz$) and $Q$.  
$Q = 0C, 0.1pC, 0.2pC$ for (a) $R_0 = 2 \mu$m and (b) $Q = 0C, 0.4pC, 0.6pC$ for $R_0 = 5 \mu$m. 
(Color Online).} 
\label{rminPsplots}
\end{figure}

The surface tension greatly influences bubble dynamics and can give rise to very distinct 
behaviours for bubbles of different ambient radii $R_0$, even when all other conditions are 
identical. In smaller bubbles, surface tension is a very dominant term. 
Looking at the $R_{min}$ vs. $P_s$ plots (Figure (\ref{rminPsplots})) for $R_0 = 2 \mu$m 
and $R_0 = 5 \mu$m, we at once see a striking difference between the two. 
For the larger, 5 micron bubble, we see that on lowering the pressure $P_s$, at a value 
corresponding to $P_s = P_{fi} = 1.12 P_0$, the $R_{min}$ curves all converge to a point. 
Lowering $P_s$ further brings the bubble 
to a cross-over regime, until reaching a lower pressure $P_s = {P_{fi}}_{true} \approx 0.9 P_0$, 
below which pressure, the curves largely show frequency - and charge-independent behavior. 
As will be seen in the ensuing paragraphs, $P_s = P_{fi}$ is actually the transition pressure 
$P_{tr}$ beyond which violent bubble collapse occurs. 

For the smaller, 2 micron bubble, we do not see any cross-over regime, and the pressure 
$P_s = P_{fi} = 1.2 P_0$ where all curves converge such that for $P_s < P_{fi}$ charge or 
frequency-dependence of the curves is suppressed, is actually less than the transition pressure 
$P_{tr} = 1.3 P_0$ for $R_0 = 2 \mu$m. 

Introduction of charge on the bubble serves to dramatically move $P_{tr}$ to a lower value, 
with the effective surface tension being reduced due to electrostatic interaction, and its 
effect being enhanced due to the bubble's smaller dimensions. 

One obvious measure of the relative extremal values of the bubble dimensions is $R_{max}/R_{min}$. 
Other measures used to quantify the bubble's dimensions are the expansion ratio 
$E \equiv R_{max}/R_0$, and the compression ratio $C \equiv R_{min}/R_0$. 
\begin{figure}
\includegraphics[height=15cm,width=7.5cm,angle=0]{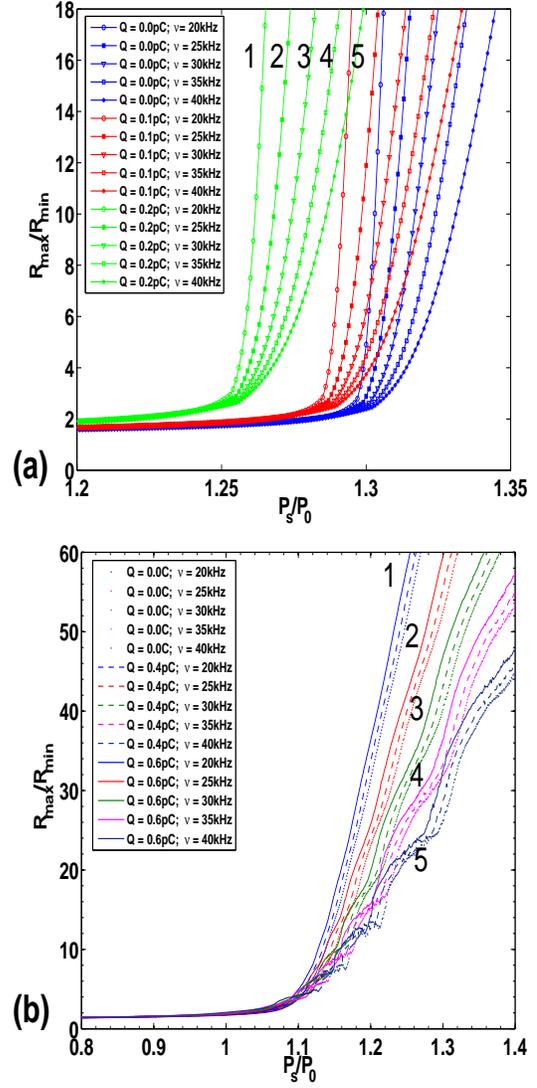}
\caption{$R_{max}/R_{min}$ vs $P_s$ for different values of $\omega$ and $Q$, for a) $R_0 = 2 \mu$m ($Q=0pC,0.1pC,0.2pC$); and b) $R_0 = 5 \mu$m ($Q=0pC,0.4pC,0.6pC$) 
($\nu= 20,25,30,35,40kHz$). Curves labelled as 1,2,3,4, \& 5 correspond to $\nu = 20,25,30,35,40kHz$, respectively. Increasing $Q$ shifts curves upwards \& to the left. (Color Online).}
\label{rmaxminplot}
\end{figure}
Figure (\ref{rmaxminplot}) shows plots of the relative extremal bubble radius measure, 
$R_{max}/R_{min}$ as a function of the amplitude $P_s$ of the driving pressure wave for two 
different bubble radii: $R_0 = 2 \mu$ m and $R_0 = 5 \mu$ m. 
We introduce yet another useful and significant measure of the relative extent of bubble 
expansion to compression, which we term the expansion-compression ratio, $\zeta$
\begin{equation}
\zeta  \equiv (E-1)/(1-C) = (R_{max} - R_0)/(R_0 - R_{min}).
\end{equation}

\begin{figure}
\includegraphics[height=17cm,width=8.5cm,angle=0]{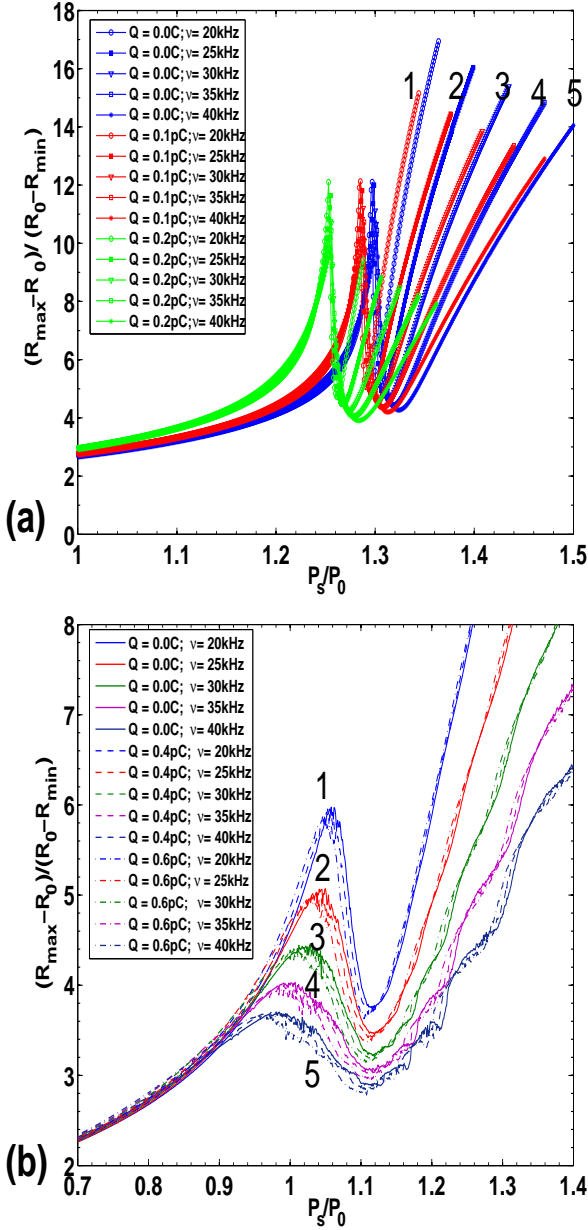}
\caption{Expansion-Contraction ratio $\zeta$ for (a) $R_0 = 2 \mu$m ($Q = 0pC,0.1pC,0.2pC$) and 
(b) $R_0 = 5 \mu$m ($Q=0pC,0.4pC,0.6pC$)as a function of $P_s$. Curves labelled as 1,2,3,4, \& 5 
correspond to $\nu = 20,25,30,35,40kHz$, respectively. Increasing $Q$ shifts curves downwards 
\& to the left. (Color Online).}
\label{ECratiofig}
\end{figure}
 Investigating the dependence of this EC ratio $\zeta$ on the amplitude of applied pressure 
$P_s$ yields some delightful results and clearly shows the great utility of this measure of 
bubble expansion/contraction. Figure (\ref{ECratiofig}) is a plot of $\zeta$ as a function of 
$P_s$ for two different ambient radius values, $R_0 = 2 \mu$m and $R_0 = 5 \mu$m. 
For bubbles with not too small ambient radius $R_0$ (for example, for $R_0 = 5 \mu$m), 
at low values of $P_s$, the expansion-compression ratio $\zeta$ becomes independent of charge 
and frequency below 
some $P_s = {P_s}_l$, and $\zeta$ curves for various frequencies and charges all superimpose 
(Figure(\ref{ECratiofig}(b))). 
This behaviour is not shown by smaller bubbles (for example, for $R_0 = 2 \mu$m), whose $\zeta$ 
curves instead show distinct charge and frequency dependence even at very low amplitudes of 
driving pressure (Figure(\ref{ECratiofig}(a))). 

\begin{figure}
\includegraphics[height=5cm,width=8.3cm,angle=0]{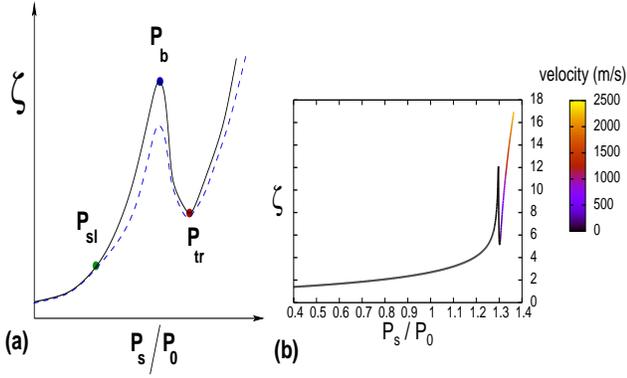}
\caption{(a).The points on the graph where frequency \& charge -independence sets in ($P_{sl}$), 
the Blake threshold ($P_b = P_{Blake}$) and the upper transient threshold pressure ($P_{tr}$) are 
shown in the schematic figure. The dashed and solid curves typify different charge and frequency 
values, which superimpose at pressures below $P_{sl}$. (b). Plot of the $\zeta$ curve with the key 
for the maximum radial velocity shown on the right, for a given frequency and charge. As can be seen, 
for pressures greater than the upper transient threshold pressure $P_{tr}$, bubble collapse occurs 
violently, at greater velocities. (Color Online).}
\label{3Dblakeplot}
\end{figure}

In general for all bubble sizes, the following generic behaviour of the EC ratio $\zeta$ is 
shown: at lower pressures, till a certain pressure $P_s =  P_{sl}$, $\zeta$ shows only very weak 
dependence on charge and driving frequency. In Figure (\ref{3Dblakeplot}(a)) the dashed and 
solid curves are shown as representative of different charge and frequency values, which 
are coincident at pressures below $P_{sl}$. 
With increasing $P_s$, the EC ratio increases to a peak at a critical pressure value $P_b$, 
followed by a short, steep dip upto a second critical pressure value $P_{tr}$. This is followed 
further by a regime of an ever-increasing  $\zeta$ with increasing $P_s$ 
(Figure (\ref{3Dblakeplot}(a))). 
Between $P_b$ and $P_{tr}$ lies a region of negative slope, which is essentially an unstable, 
transient region.\\
Surface tension is overcome at a critical radius corresponding to the pressure $P_b$ after 
which significant bubble expansion occurs and the motion is transient until pressure $P_{tr}$. 
For $R_0$ being sufficiently large, bubble collapse can not be completed fully during the 
compression part of the cycle of applied pressure and bubble motion is then stable. This explanation 
accounts for the presence of two thresholds, enclosing a transient regime with stable regions on 
either side~\cite{neppiras}.
At the lower threshold, it can be seen from Figures (\ref{ECratiofig}, \ref{3Dblakeplot}) the 
transition from stable to transient conditions occurs very steeply.\\
The lower transient threshold pressure $P_b$  delineates a pressure value above which the bubble 
expansion occurs dramatically, and this in fact equals the Blake threshold pressure, $P_b = P_{Blake}$.\\ 

It is important to recall at this 
point that while the Blake threshold is a measure of the onset of rapid bubble expansion, it gives 
us no information at all about bubble implosion~\cite{prosperettiUltra}. The upper transient threshold 
pressure, $P_{tr}$, on the other hand, is the value of $P_s$ above which violent bubble contraction 
begins. It can be seen, by inspecting the $R_{min}$ versus $P_s$ curve (Figure \ref{rminPsplots}) 
that at $P_s = P_{tr}$, $R_{min} = R_0/2$. This, in fact, identifies the transition to a strong 
collapse regime  from weaker oscillations \cite {hilgenfeldt}.
This can be verified by incorporating the radial velocity into the EC plot. As can be seen in 
Figure (\ref{3Dblakeplot}(b)), where a scale representing the magnitude of maximum radial velocity has 
been included, it is only at driving pressures above the upper transient threshold $P_{tr}$ that 
velocities rise dramatically to high values, where as at lower $P_s$, the radial oscillations 
occur more slowly. 

\begin{figure}
\includegraphics[height=13cm,width=7cm,angle=0]{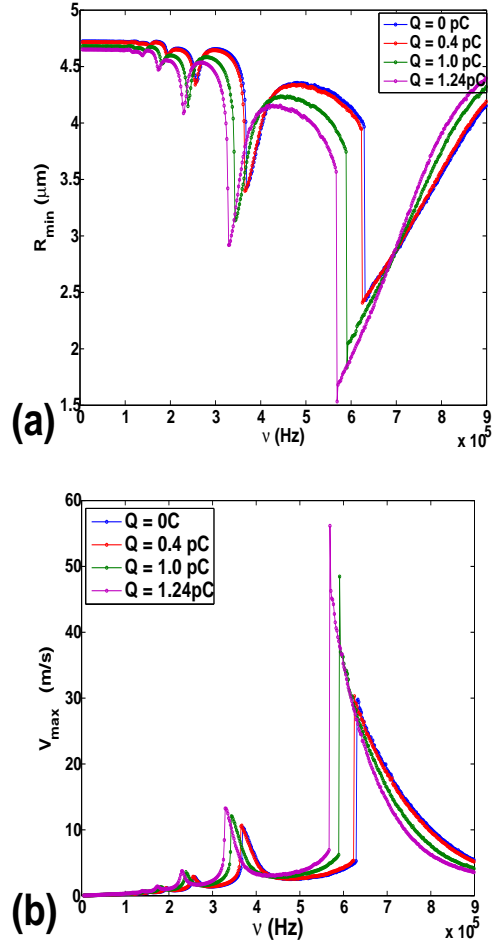}
\caption{Frequency response curves for minimum radius (a) and maximum velocity (b) of the bubble. 
$R_0=5\mu$m, $Q=0pC, 0.4pC,1.0pC,1.24pC$. (Color Online).}
\label{rmin-vel_responsecurves}
\end{figure}

In Figure (\ref{rmin-vel_responsecurves})(a) and (b) we show the frequency response diagrams for 
the minimum radius and maximum velocity attained by a 5 micron bubble in water at the low driving 
pressure of $0.4 P_0$ and carrying charges $Q= 0, 0.4, 1.0$ and $1.24$pC. 
It is clear that increasing the magnitude of charge present on the bubble increases the magnitude 
of the response and advances it to lower frequencies. 
The peaks in the response diagram appear much earlier, at lower frequencies, for higher charges 
and their enhanced magnitudes (smaller minimum radius and larger maximal velocity) point to more 
violent collapse.\\
We can make a further, very rough estimate of the maximal bubble expansion limits at this transient 
pressure. We first consider the following vastly simplifying assumptions: that at 
$R \rightarrow R_{min}$, ${\dot R} \rightarrow c$, where $c$ is the speed of sound in the liquid. 
We further assume, for purpose of this estimation,  that as $R \rightarrow R_{max}$, ${\dot R} \rightarrow 0$.\\  At its local minimum at $P_s = P_{tr}$, $\zeta=\zeta^{tr}$ will 
satisfy $({\frac{\partial \zeta}{\partial P_s}})|_{P_s = P_{tr}} = 0$, so that we get 
\begin{equation}
\frac{\partial R_{max}}{\partial P_s} = -\frac{R_{max} - R_0}{R_0 - R_{min}}\frac{\partial R_{min}}{\partial P_s} = -\zeta^{tr}\frac{\partial R_{min}}{\partial P_s}.
\label{ratioeqn1}
\end{equation}
Now using 
\begin{eqnarray}
\frac{\partial R_{min}}{\partial P_s} &\approx& \bigg(-5(P_a + \frac{2\sigma}{R_0} - \frac{Q^2}{8 \pi \epsilon R_0^4})\frac{R_0^5}{R_{min}^6} + \frac{2\sigma}{R_{min}^2} \nonumber \\ 
&-& \frac{4Q^2}{8 \pi \epsilon R_{min}^5} + 4 \eta \frac{c}{R_{min}^2}\bigg)^{-1} \nonumber \\
\frac{\partial R_{max}}{\partial P_s} &\approx& \bigg(-5(P_a + \frac{2\sigma}{R_0} - \frac{Q^2}{8 \pi \epsilon R_0^4})
\frac{R_0^5}{R_{max}^6} + \frac{2\sigma}{R_{max}^2} \nonumber \\
&-& \frac{4Q^2}{8 \pi \epsilon R_{max}^5}\bigg)^{-1},
\end{eqnarray}
in Equation (\ref{ratioeqn1}), we get
\begin{eqnarray}
\zeta^{tr} &=& -\left(\frac{5\phi}{R_{max}^6} - \frac{2\sigma}{R_{max}^2} 
+ \frac{4 Q^2}{8 \pi \epsilon R_{max}^5}\right) \Big/\nonumber \\
&&\left(\frac{5\phi}{R_{min}^6} - \frac{2\sigma}{R_{min}^2} + \frac{4 Q^2}{8 \pi \epsilon R_{min}^5} 
- \frac{4\eta c}{R_{min}^2}\right),
\end{eqnarray}
where $\phi/R_0^5$ denotes the equilibrium pressure of the gas in the bubble (Eqn.(6)). 

If we further consider the extremal case of $R_{min} \approx h$, this becomes 
\begin{equation}
\zeta^{tr} = -\frac{\left(\frac{5\phi}{R_{max}^6} - \frac{2\sigma}{R_{max}^2} 
+ \frac{4 Q^2}{8 \pi \epsilon R_{max}^5}\right)}{\left(\frac{5\phi}{h^6} 
- \frac{2\sigma}{h^2} + \frac{4 Q^2}{8 \pi \epsilon h^5} - \frac{4\eta c}{h^2}\right)}.
\label{ratioheqn}
\end{equation} 
Now extremizing equation (\ref{ratioheqn}) with respect to $R_{max}$ yields a quartic equation 
for $R_{max} = R_{max}^{tr}$ at $P_s = P_{tr}$ 
\begin{equation}
R_{max}^4 - \frac{5 Q^2}{8 \pi \epsilon \sigma}R_{max} - \frac{15 \phi}{2 \sigma} = 0.
\end{equation}
In the uncharged case, $Q=0$, this simplifies to 
\begin{equation}
R_{max}^{tr} = \left(\frac{15 \phi}{2 \sigma}\right)^{1/4},
\label{RmaxQ0eqn}
\end{equation}
which is the value of $R_{max}$ at the transition point at $P_s = P_{tr}$, made under all the 
simplifying assumptions mentioned above. It will be noted that for mid-sized microbubbles, for 
example for $R_0 = 5 \mu$m (plots for which have been shown), the point of transition $P_{tr}$ 
is approximately a constant and is largely independent of the driving frequency or the charge.\\  

A comparison of the estimate of $R_{max}$ so obtained from the above equation for the uncharged case 
to the value obtained numerically is given in Table 1 below. Though the estimated values are far 
from accurate in many cases, they do, nonetheless, give a quick and useful estimation of $R_{max}$ 
at a driving pressure equalling the upper transient threshold of pressure $P_{tr}$. 

\begin{table}
\caption{$R_{max}^{tr}$ obtained from Eqn.(32) \& graphically.} 

\begin{tabular}{|l|l|l|} 
\hline
\multicolumn{3}{|c|}{Table 1}\\
\hline
$R_0$  & $R_{max}^{tr}$ (eqn.) & $R_{max}^{tr}$ (graph)\\
\hline
2 $\mu$m &4.87 $\mu$m &7.14 $\mu$m\\
3 $\mu$m &7.78 $\mu$m &10.65 $\mu$m\\
4 $\mu$m &10.92 $\mu$m &12.64 $\mu$m\\
5 $\mu$m &14.23 $\mu$m &14.3 $\mu$m\\
7 $\mu$m &21.3 $\mu$m &17.1 $\mu$m\\
\hline
\end{tabular}
\end{table}

These estimates though rather crude, might provide useful measures for avoiding undesirable 
regimes involving violent bubble collapses in medical diagnostics and applications. 

\section{Conclusions}
In this work we have investigated the dynamics of a bubble forced by an ultrasound field, 
and seen how charge influences its behaviour. Our calculations are for a system where an 
adiabatic equation of state prevails, with $\Gamma = 5/3$. We make several interesting 
observations. Charge serves to reduce the effective surface tension of the bubble. This 
causes a charged bubble to not only expand to a larger radius as compared to a neutral 
bubble but also collapse to a smaller minimum radius, the lower bound of which is given 
by the van der Waals hard core radius. The charged bubble's collapse is also more violent, 
with radial velocities being reached being greater. Charge influences and modifies the 
liquid pressure profile. The effects are more marked for bubbles of smaller dimensions, 
where surface tension has a predominant influence. 
Studies of the effect of the amplitude of the forcing pressure wave show that introduction 
of charge serves to lower the Blake threshold so that the transition to violent collapses 
and oscillations occur at lower levels of pressure, especially for microbubbles of smaller 
dimensions and submicron bubbles. We have obtained expressions for the Blake threshold 
in the presence of charge.  

We introduce a measure of the extremal dimensions reached by a bubble, which is a  
quantity $\zeta = (E-1)/(1-C)$ where $E$ and $C$ are the expansion and compression ratios, 
respectively. The advantage of plotting $\zeta$ as a function of $P_s$ is that 
it captures the distinct positions and behaviors of both the Blake threshold pressure 
as well as the upper transient threshold pressure, between which points lies a regime 
of instability for the bubble.

The maximum magnitude of charge a bubble can carry in the system tends to converge to a 
single asymptotic value, $Q_{max}$, and likewise the minimum radius converges to a 
single value for all frequencies, which is however modulated by the radial length scale 
cut-off provided by the van der Waals hard core radius.
Exceeding the magnitude of charge $Q_h$ causes the bubble to contract to values of $R_{min}$ 
that are less than $h$ which cannot be physically reached and hence provide a physical 
upper bound for the system's charge. 
We also investigated Rayleigh collapse for the bubble, obtaining an expression for the 
characteristic Rayleigh radius $R_r$. Frequency-dependence of the minimum radius is 
also captured. We also obtain approximate relations for the maximal radius at the 
critical transition point. 
Most of the numerical results presented in this work are for high pressures, corresponding 
to regimes of violent bubble collapse. We have demonstrated the importance of including 
charge in investigating the expansion and contraction of a bubble under forcing. We have 
also found scaling relations for extremal radial dimensions. Other results, including an 
investigation of the bifurcation structure as a function of charge, scaling relations for 
the maximal charge, etc. are being reported elsewhere~\cite{ashok}. 

\section*{Acknowledgments} T.H. is supported by a Rajiv Gandhi National Fellowship from the 
University Grants Commission, New Delhi, for his doctoral studies. 

\end{document}